\documentclass[a4paper,10pt,oncolumn,twoside]{cpc-hepnp}
\usepackage{upgreek,fancyhdr}
\usepackage{xcolor}
\usepackage{multicol}
\usepackage{multirow}
\usepackage{graphicx}
\usepackage{booktabs}
\usepackage{amssymb,bm,mathrsfs,bbm,amscd}
\usepackage[tbtags]{amsmath}
\usepackage{lastpage}
\usepackage[english]{babel}
\usepackage{epsfig}
\usepackage{graphicx}
\usepackage{epstopdf}
\usepackage{graphbox}

\begin{document}

\fancyhead[c]{\small Chinese Physics C~~~Vol. xx, No. x (2023) xxxxxx}
\fancyfoot[C]{\small 010201-\thepage}
\footnotetext[0]{Received \today}

\title{Inferring the spin distribution of binary black holes using deep learning}

\author{Li Tang$^{1;1)}$\email{tang@mtc.edu.cn}
\quad Xi-Long Fan$^{2}$}

\maketitle
\hspace{1cm}

\address{$^1$ School of Math and Physics, Mianyang Teachers' College, Mianyang 621000, China\\
$^2$ School of Physics and Technology, Wuhan University, Wuhan, Hubei 430072 China}

\begin{abstract}
  The spin characteristics of black holes offer valuable insights into the evolutionary pathways of their progenitor stars, crucial for understanding the broader population properties of black holes. Traditional Hierarchical Bayesian inference techniques employed to discern these properties often entail substantial time investments, and consensus regarding the spin distribution of Binary Black Hole (BBH) systems remains elusive. In this study, leveraging observations from GWTC-3, we adopt a machine learning approach to infer the spin distribution of black holes within BBH systems. Specifically, we develop a Deep Neural Network (DNN) and train it using data generated from a Beta distribution. Our training strategy, involving the segregation of data into 10 bins, not only expedites model training but also enhances the DNN's versatility and adaptability to accommodate the burgeoning volume of gravitational wave observations. Utilizing Monte Carlo-bootstrap (MC-bootstrap) to generate observation-simulated samples, we derive spin distribution parameters: $\alpha=1.3^{+0.25}_{-0.18},\beta=1.70^{+0.24}_{-0.29}$ for the larger BH sample and $\alpha=1.37^{+0.31}_{-0.20},\beta=1.63^{+0.30}_{-0.20}$ for the smaller BH sample. Within our constraints, the distributions of component spin magnitudes suggest the likelihood of both black holes in the BBH merger possessing non-zero spin.
\end{abstract}


%

\footnotetext[0]{\hspace*{-3mm}\raisebox{0.3ex}{$\scriptstyle\copyright$}2019
Chinese Physical Society and the Institute of High Energy Physics
of the Chinese Academy of Sciences and the Institute
of Modern Physics of the Chinese Academy of Sciences and IOP Publishing Ltd}%

\section{Introduction}\label{sec:introduction}

Gravitational waves arising from the merger of binary systems, were initially detected directly by Advanced LIGO in 2015 \cite{LIGOScientific:2016aoc}, marking a seminal milestone in gravitational wave research. These waves encapsulate astronomical information that is not accessible through electromagnetic signals and inaugurate the era of gravitational-wave astronomy. To date, Advanced LIGO and Advanced Virgo, the ground-based gravitational wave detectors, have successfully completed their third operational phase, observing nearly 100 GW events associated with the merger of compact binary systems \cite{LIGOScientific:2018mvr,LIGOScientific:2020ibl,LIGOScientific:2021usb,KAGRA:2021vkt}. Among these events, there were two mergers involving neutron star-black hole binaries (NSBH) and two involving binary neutron stars (BNS). Focusing primarily on binary black hole (BBH) events, the analysis of BBH parameter distributions contributes to our understanding of the massive stellar evolution and binary formation processes \cite{Roulet:2021hcu,Galaudage:2021rkt,Mould:2022xeu}. Various hypotheses have been postulated to elucidate the formation of compact binaries. One of the most promising theories involves the isolated evolution of binaries in the galactic field, where mechanisms like common-envelope phases \cite{Nelemans:2001hp,Belczynski:2014iua,Giacobbo:2018etu} and chemically homogeneous evolution \cite{Mandel:2015qlu,deMink:2016vkw} play pivotal roles. A contrasting theory, contending for prominence, posits dynamical assembly within densely populated stellar environments like globular clusters \cite{Downing:2009ag,Rodriguez:2016kxx}, nuclear star clusters \cite{Antonini:2016gqe,Petrovich:2017otm}, and young stellar clusters \cite{Ziosi:2014sra,Chatterjee:2016thb}.

The spin of BH play a pivotal role not only in analyzing the evolutionary processes of BH but also in elucidating the hierarchical assembly of multiple generations of stellar-mass BH mergers. BHs resulting from stellar collapse typically exhibit negligible spins, as the expulsion of matter dissipates angular momentum during the collapse process \cite{Fuller:2019sxi,Ma:2019cpr}. Conversely, BHs formed through multiple generation merger may manifest non-negligible spins. Abbott et al. \cite{LIGOScientific:2020kqk, KAGRA:2021duu} inferred that BHs demonstrate non-zero spins, with notable misalignments concerning the binary orbital angular momentum. Similar findings are corroborated in these references \cite{Callister:2022qwb,Mould:2022xeu,Tong:2022iws}. Galaudage et al. \cite{Galaudage:2021rkt} conducted an analysis of BBH merger data from the second LIGO-Virgo gravitational-wave transient catalog (GWTC-2) and identified two distinct subpopulations, where a significant proportion of BHs exhibited negligible spin, contrasting with a smaller fraction demonstrating non-negligible spin. Additionally, Adamcewicz et al. \cite{Adamcewicz:2023szp} observed, within data from GWTC-3, that the both-spin framework, wherein both black holes exhibit apparent spins, outperforms the single-spin framework.

Conclusions derived from diverse methodologies manifest discrepancies. In addition to the aforementioned application of conventional methods for inferring population parameters, there is an emerging trend involving the integration of machine learning techniques for this purpose. Wong et al. \cite{Wong:2019uni} exemplify this trend by training a machine learning emulator on numerical population synthesis predictions, derived from simulations of isolated binary stars. Subsequently, the emulator is incorporated into a Bayesian hierarchical framework to quantify the natal kicks received by BHs at birth. Following a similar trajectory, Mould et al. \cite{Mould:2022ccw} utilized fully connected deep neural networks (DNN) to reconstruct probability density functions. By training these networks with simulations employing various power-law parametrizations and combining with hierarchical Bayesian inference, they constrained the properties of BH mergers using data from GWTC-3. Assuming a spin distribution for first-generation BHs characterized by isotropic direction and uniform magnitude, parameterized by the maximum spin $\chi_{\rm{max}}$, Mould et al. \cite{Mould:2022ccw} determined that the maximum spin is $\chi_{\rm{max}}=0.39^{+0.08}_{-0.07}$. The distribution of effective aligned spins exhibited $|\chi_{\rm{eff}}|<0.46^{+0.04}_{-0.06}$, and the effective precessing spins displayed multimodal characteristics.

Although the traditional Markov Chain Monte Carlo (MCMC) method, rooted in Bayesian analysis, has become a popular method used in the GW population analysis, MCMC typically incurs high computational costs, particularly in high-dimensional parameter spaces or complex models. Achieving convergence demands a substantial number of samples, leading to extended computation times. Additionally, MCMC results are also influenced by the choice of prior distributions. Recently, machine learning methods, especially deep learning (DL), excel in rapidly processing large-scale datasets and high-dimensional problems through parallel computation and efficient optimization algorithms. DL techniques autonomously learn intricate patterns and features from data, thereby diminishing reliance on prior knowledge. They achieve robust performance through training on extensive datasets, offering a compelling alternative to traditional MCMC methods in various applications. In this work, we will utilize deep learning methodology to investigate the spin distribution within BBH merger data from the GWTC-3 catalog, as outlined in Ref.\cite{KAGRA:2021duu}. Unlike previous works \cite{Wong:2019uni, Mould:2022ccw} where networks were designed to act as a density estimator of a population model, our network is specifically trained to estimate the parameters of a spin model. Consequently, our network architecture is more straightforward and benefits from reduced training times, as it is trained exclusively on spin simulations generated from the Beta model. The model parameters of the spin observation can be extracted by inputting the observational data into the well-trained network.

The following segments of this paper are structured as outlined below: In Section 2, we provide an introduction to both the data and the spin model. Additionally, we elaborate on the network utilized to express the distribution of observations. Section 3 is dedicated to the presentation of results, accompanied by a thorough examination of our network through simulations. The discussion and conclusions are expounded upon in Section 4.

\section{Data and methodology}\label{sec:data}

\begin{figure*}[htbp]
 \centering
 \includegraphics[width=0.95\textwidth]{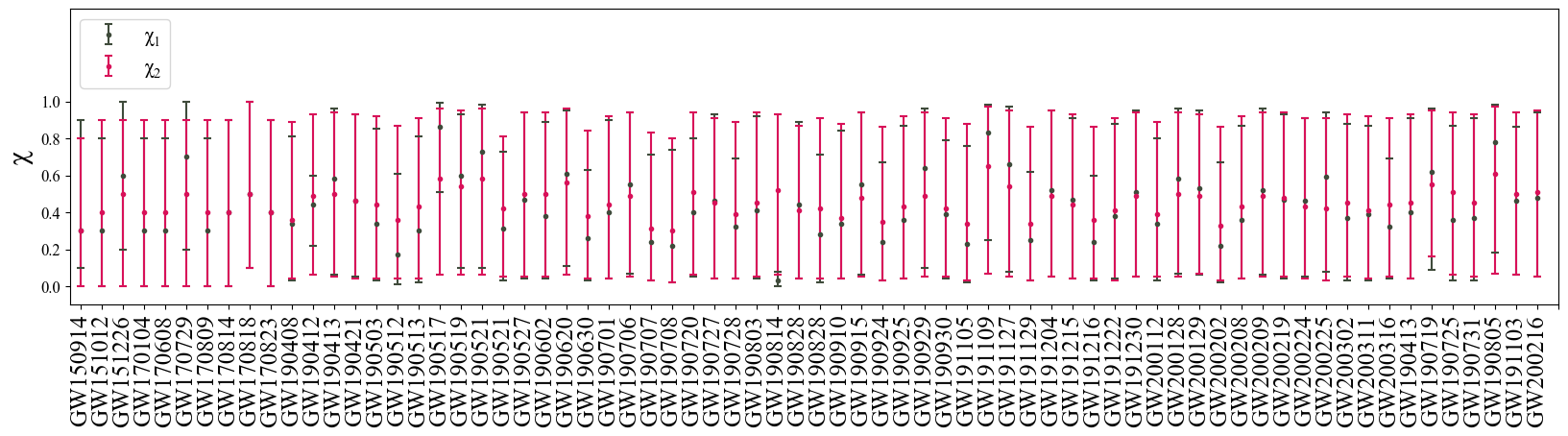}
 \caption{Spin magnitudes of BBH events utilized in this paper extracted from Ref.\cite{KAGRA:2021duu},denoted as $\chi_1$ and $\chi_2$ for the spin magnitudes of the first and second black holes, respectively.}\label{fig:GW_event}
\end{figure*}

Deep learning utilizes artificial neural networks (ANNs) as an approximate function to characterize the relationship between the input data and output data, where ANNs typically consist of an input layer to receive input, an output layer to produce results, and one or more hidden layers between them. When there are more than two hidden layers, the neural network is considered a deep neural network (DNN). Deep learning has been widely applied in the fields of cosmology \cite{Escamilla-Rivera:2019hqt,Luongo:2020hyk,Tang:2020nmw,Liu:2023rrr} and gravitational wave astronomy research \cite{Wang:2019zaj,Luo:2019hvt,Gerosa:2020pgy,Chen:2020ehw,Wang:2021srv,Ma:2022esx,Wang:2023lif,Tang:2024hvk}. In this work, we construct a fully connected deep neural network to investigate the spin magnitude distribution of BBH through the Gravitational-Wave Transient Catalog 3 (GWTC-3) \cite{KAGRA:2021duu}. Excluding events involving neutron stars, we obtain a total of 70 BBH merger events for the study of population information. For visualization purposes, we depict the spin magnitudes of the available BBH events in Figure \ref{fig:GW_event}.

Our methodology is grounded in the spin magnitude model, wherein each component spin magnitude of BBH mergers is delineated by a Beta distribution \cite{KAGRA:2021duu, Wysocki:2018mpo}:
\begin{equation}\label{Eq:spin_model}
p(\chi|\alpha,\beta)\propto \chi^{1-\alpha}(1-\chi)^{1-\beta},
\end{equation}
Here, $\chi$ signifies the magnitude spin of one of the black holes in the BBH merger and is constrained within the interval $\chi\in[0,1]$. The parameters $\alpha$ and $\beta$ function as normalization constants, adhering to the stipulations $\alpha\geq1$ and $\beta\geq1$ to ensure a non-singular distribution. We utilize simulated data generated from the aforementioned model to facilitate network training. Although the direct utilization of the 70 spin magnitude observation could suffice for obtaining distribution parameters, optimizing training efficiency through data preprocessing is essential. Consequently, we segment the 70 spin magnitudes into 10 bins and employ the normalized counts of each bin as training features. In essence, the probability density values serve as features, while the corresponding parameters act as labels.

During the preparation of training data, for the purpose of precision enhancement, we directly generate training data, i.e., probability density samples, from the distribution model (i.e., Eq.(\ref{Eq:spin_model})), based on a series of parameters denoted by $\boldsymbol{y}_i=\{\alpha,\beta\}_i$ (the training labels), where $\alpha\in[1.2,8]$ and $\beta\in[1.2,8]$, both with an interval of 0.2, resulting in a total of 35$\times$35 parameter combinations. For the $i$th set of parameters, the corresponding ten probability density values (the training features) are represented by a vector $\boldsymbol{x}_i$.

\begin{figure*}[htbp]
 \centering
 \includegraphics[width=0.8\textwidth]{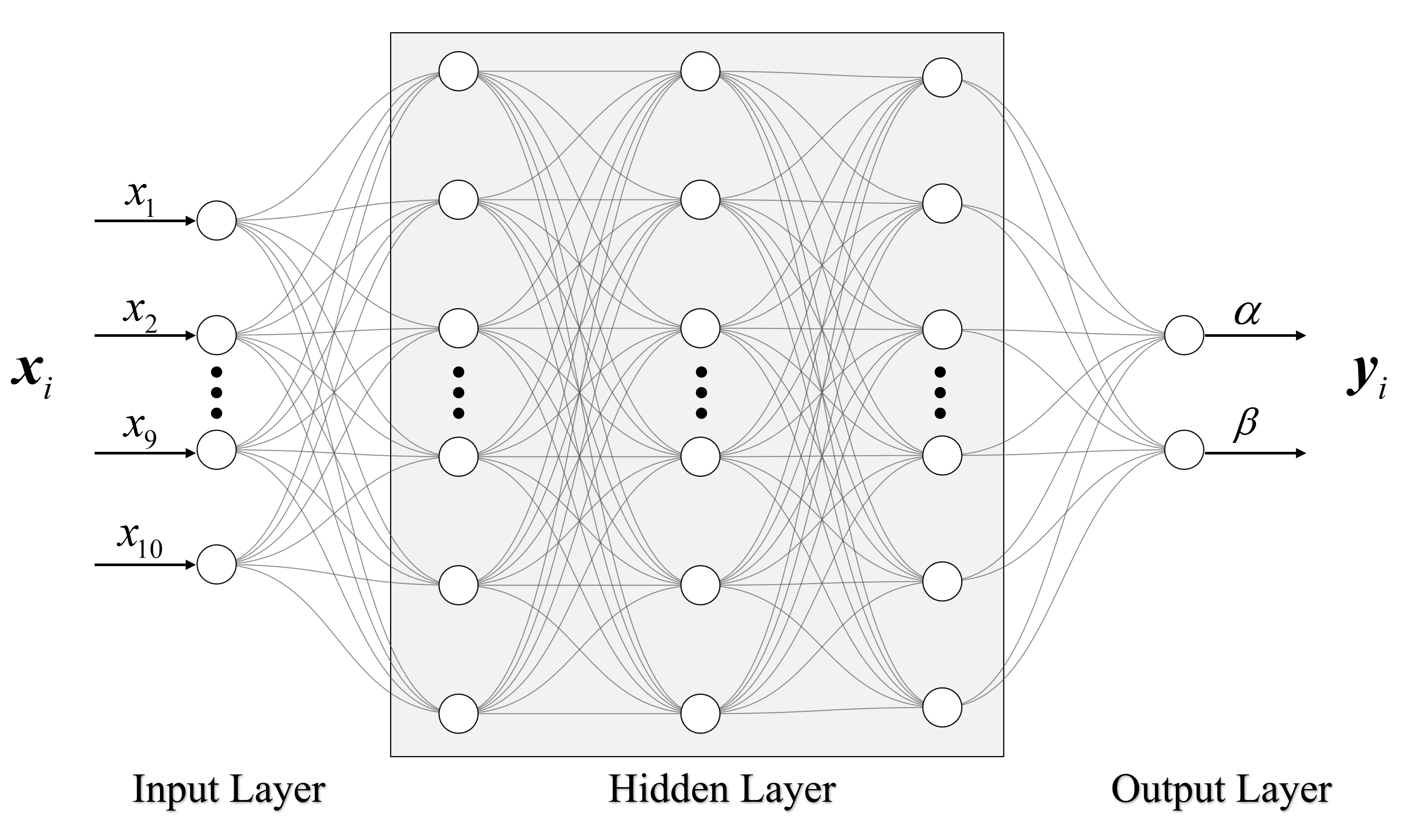}
 \caption{The architecture of DNN with three hidden layers. The input and output are the probability density values and the corresponding distribution parameters, respectively. The number of circles represents the number of neurons in each layer.}\label{fig:DNN}
\end{figure*}

The structure of our DNN, comprising three hidden layers, is depicted in Figure \ref{fig:DNN}. The initial layer, i.e. input layer, comprising ten neurons, accepts the probability density vector, while the final layer, i.e. output layer, housing two neurons, outputs the parameter vector. Intermediary between these layers are three hidden layers, the first with 200 neurons, the second with 400 neurons, and the third with 200 neurons. These layers facilitate the transformation of information from preceding layers to succeeding layers through a combination of linear transformations and nonlinear activations, the latter being instrumental in augmenting network performance.

The matrix algorithms employed within our DNN are delineated as follows:

1)  Initialization (for $l$=1):
\begin{equation}
\boldsymbol{A}^l=\boldsymbol{X},
\end{equation}

2) Hidden layers (for $l$=2 to 4):
\begin{equation}
\boldsymbol{A}^{l}=\sigma\left(\boldsymbol{A}^{l-1}\boldsymbol{W}^{l}+\boldsymbol{B}^{l}\right),
\end{equation}

3) Output (for $l$=5):
\begin{equation}
\boldsymbol{ \hat{Y} }=\boldsymbol{A}^{l-1}\boldsymbol{W}^{l}+\boldsymbol{B}^{l},
\end{equation}
here, the input matrix $\boldsymbol{X}$ possesses dimensions $m\times n$, where $m$ represents the size of the training set and $n$ signifies the number of features. $\boldsymbol{A}^{l}$ denotes the output of the $l$th layer, characterized by dimensions $m\times k$, where $k$ corresponds to the number of neurons in the $l$th layer. $\boldsymbol{W}^{l}$ stands for the weight matrix connecting the $(l-1)$th layer to the $l$th layer, with dimensions $j\times k$, where $j$ denotes the number of neurons in the $(l-1)$th layer. $\boldsymbol{B}^{l}$ represents the bias matrix, shaped $m\times 1$. $\sigma$ denotes the activation function, specified as the Rectified Linear Unit (ReLU) function:
\begin{equation}
\sigma(x)=
\begin{cases}
0,& x\leq0\\
x,& x>0
\end{cases}
\end{equation}
$\boldsymbol{\hat{Y}}$ constitutes the DNN's prediction, with dimensions $m\times p$, where $p$ designates the number of labels. It is pertinent to note that computations within the input and output layers are linear. We employ the Mean Squared Error (MSE) function as the loss function to quantify the difference between the prediction $\boldsymbol{\hat{Y}}$ and the labels $\boldsymbol{Y}$, leveraging the Adam optimizer to minimize this loss. The learning rate and training epoch are set to 0.001 and 5000, respectively. The ratio of the training set to the validation set is maintained at 8:2. Through training the DNN with data, the hyperparameters $\boldsymbol{W}$ and $\boldsymbol{B}$ in each layer are determined, culminating in a well-trained DNN capable of approximating a function $\boldsymbol{Y}=f_{\boldsymbol{W},\boldsymbol{B}}\left(\boldsymbol{X}\right)$. Consequently, the parameters of the spin magnitude distribution $\{\alpha,\beta\}$ can be predicted by inputting the spin observation distribution.

\begin{figure*}[htbp]
 \centering
 \includegraphics[width=0.5\textwidth]{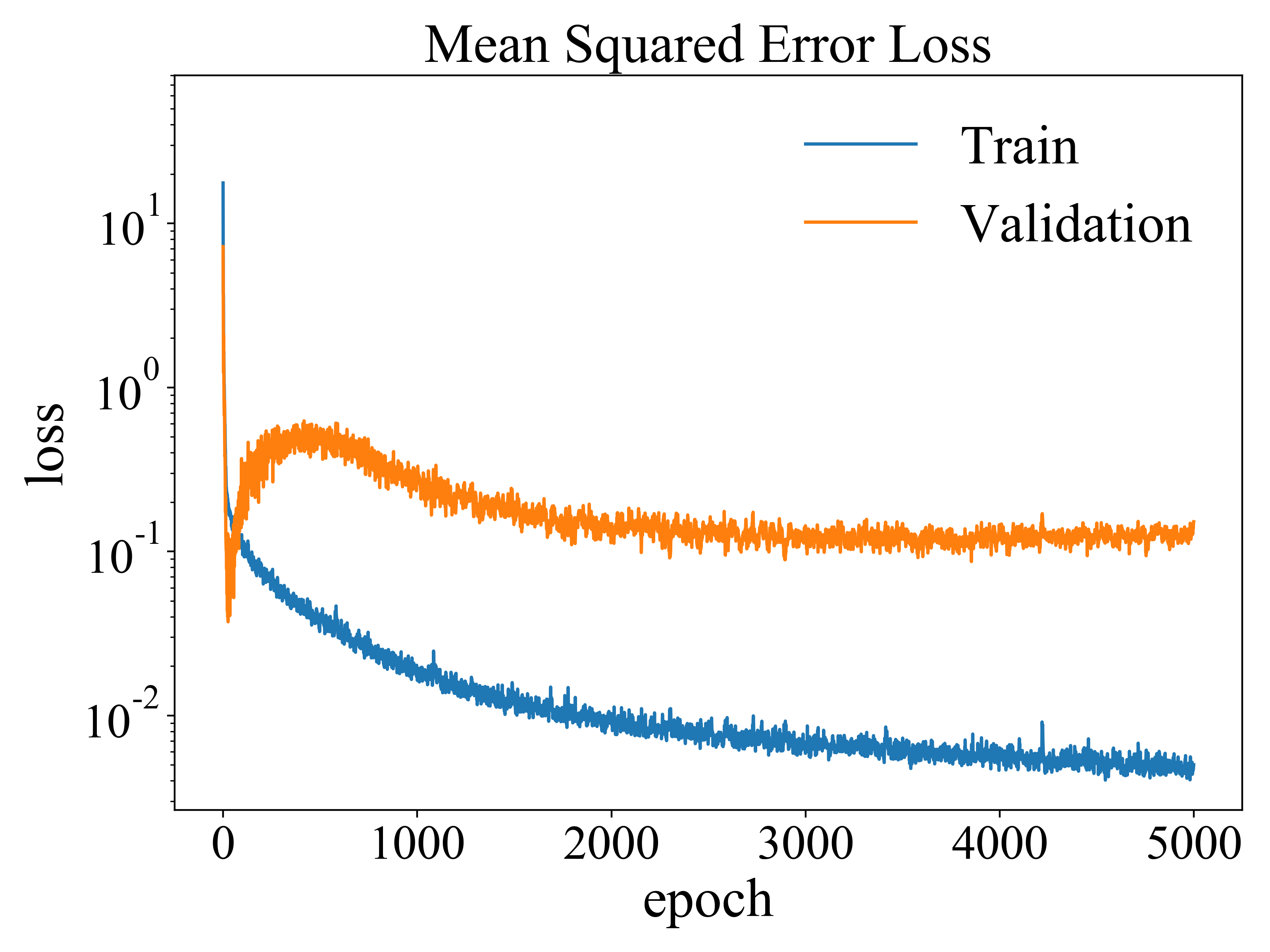}
 \caption{The Mean Squared Error (MSE) loss over the training epochs for both the train (blue) and validation (orange) sets.}\label{fig:loss of training}
\end{figure*}

The DNN is trained using the TensorFlow package\footnote{https://www.tensorflow.org}. The MSE loss throughout the training epochs for both the training and validation sets are visualized in Figure \ref{fig:loss of training}, where the MSE loss for the training set approaches 10$^{-2}$ around epoch 2000, while the MSE loss for the validation set also reaches its minimum near this epoch. This convergence suggests that our model's performance meets the practical requirements. To assess the predictive accuracy of the well-trained DNN further, a novel test set is generated for prediction evaluation. The examination reveals that the variance between the actual values and predictions does not exceed 0.02.

\section{Results and Simulations}\label{sec:results}

\begin{figure*}[htbp]
 \centering
 \includegraphics[width=0.45\textwidth]{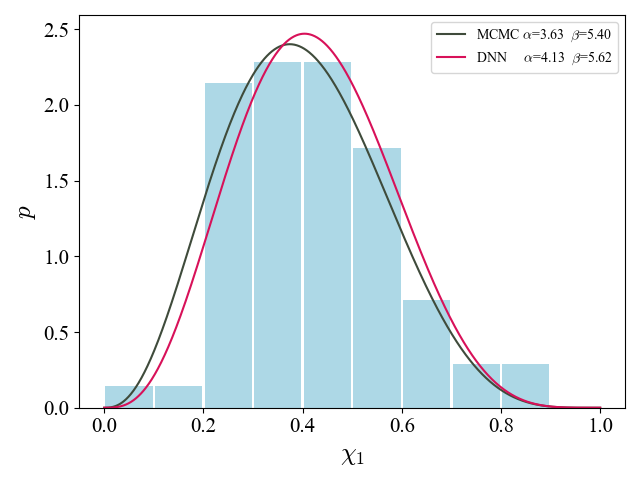}
 \includegraphics[width=0.45\textwidth]{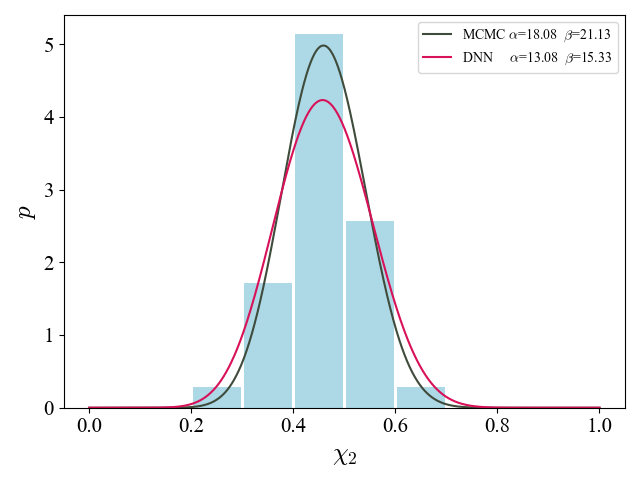}
 \caption{The histograms of spin observations: left: $\chi_1$, right: $\chi_2$. The red and olive green lines represent the predictions of DNN and MCMC methods, respectively.}\label{fig:real_spin}
\end{figure*}

Utilizing the well-trained DNN, we can infer the distribution parameters of a dataset with an unrestricted number of BBH events, benefiting from segmenting the data into 10 bins. Notably, within the spin sample of the largest BH in GWTC-3, denoted as $\chi_1$, the parameters manifest as $\alpha=4.13$ and $\beta=5.62$. Similarly, within the spin sample of the smaller BH in GW events, denoted as $\chi_2$, the parameters are determined as $\alpha=13.08$ and $\beta=15.33$. For comparative analysis, we utiliz the MCMC approach to estimate the distribution parameters. Assuming uniform priors for both $\alpha$ and $\beta$ within the interval [1, 30], the parameters are constrained $\alpha = 3.63^{+0.55}_{-0.51}$ and $\beta = 5.40^{+0.80}_{-0.73}$ for $\chi_1$, and $\alpha = 18.08^{+1.99}_{-1.82}$ and $\beta = 21.13^{+2.44}_{-2.20}$ for $\chi_2$. The results for $\chi_1$ align with those obtained from the DNN within the 1$\sigma$ confidence level, whereas the results for $\chi_2$ exhibit a deviation from the DNN outcomes. The corresponding probability density functions (PDFs) are illustrated in Figure \ref{fig:real_spin}, where the MCMC method assigns greater significance to bins with a higher number of spins compared to the DNN, which considers the distribution in its entirety. Regarding computational efficiency, the MCMC method exhibited significantly higher time consumption compared to the DL method. Notably, the uncertainties of parameters can not be provided with one sample using DL method. Furthermore, it is also not suitable to determine the parameters only with the central values of spin sample.

To address these limitations, a Monte Carlo-bootstrap (MC-bootstrap) method \cite{Bengaly:2022cgs} is proposed. The procedural steps entail: 1) Sampling the spin magnitude from the posterior distributions for the larger and smaller BHs within a specific GW event to generate samples akin to observations, denoted as sample of $\chi_1$ and sample of $\chi_2$, respectively. 2) Segregating each sample into 10 bins and normalizing the count within each bin. 3) Utilizing the normalized counts of each sample as input to the well-trained DNN to deduce the corresponding parameters $\{\alpha,\beta\}$. 4) Iterating through the aforementioned steps 1000 times to obtain parameter chains.

\begin{figure*}[htbp]
 \centering
 \includegraphics[width=0.45\textwidth]{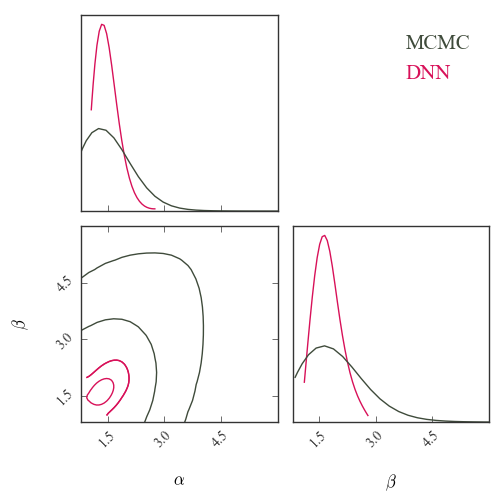}
  \includegraphics[width=0.45\textwidth]{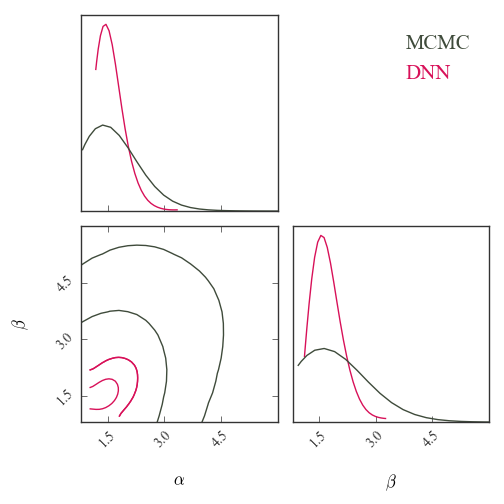}
 \caption{The 2-dimensional confidence contours and 1-dimensional PDFs for for two samples: left:$\chi_1$,right:$\chi_2$.}\label{fig:contour_real}
\end{figure*}

\begin{figure*}[htbp]
 \centering
 \includegraphics[width=0.45\textwidth]{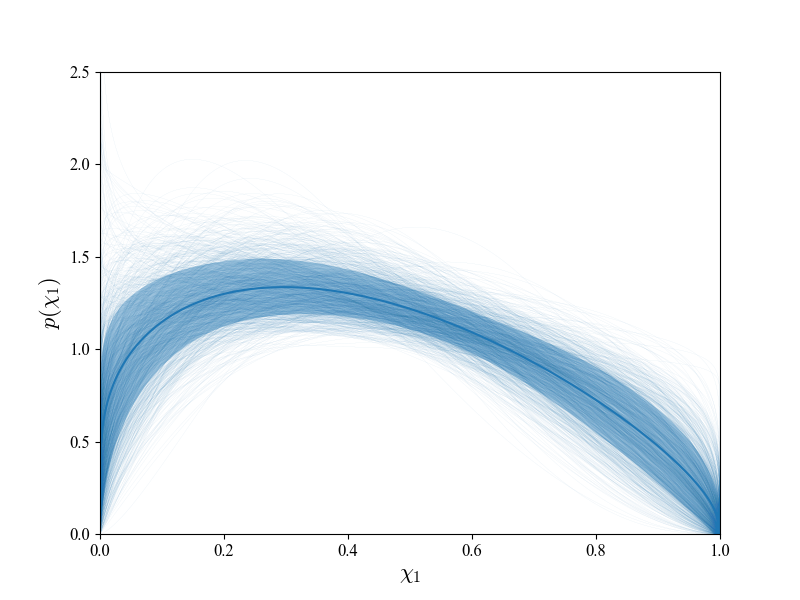}
  \includegraphics[width=0.45\textwidth]{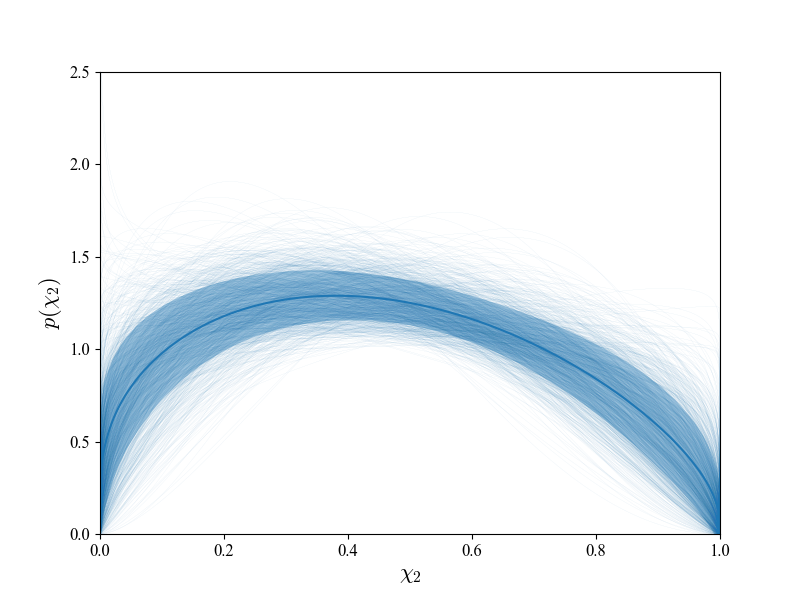}
 \caption{The distribution of component spin magnitudes: left:$\chi_1$,right:$\chi_2$. The solid blue lines and shaded regions denote the medians and central 1$\sigma$ credible bounds, respectively.}\label{fig:contour_real_bands}
\end{figure*}

The 2-dimensional confidence contours and 1-dimensional PDFs for the two samples are depicted in Figure \ref{fig:contour_real}. Within the sample of $\chi_1$, the parameters are predicted as $\alpha=1.38^{+0.34}_{-0.21}$ and $\beta=1.64^{+0.38}_{-0.26}$. Within the sample of $\chi_2$, parameters are $\alpha=1.37^{+0.31}_{-0.20}$ and $\beta=1.63^{+0.30}_{-0.20}$. The parameters are also estimated via the MCMC method, utilizing MC samples. This approach yields parameter estimates of $\alpha=1.27^{+0.44}_{-0.28}$ and $\beta=1.62^{+0.71}_{-0.45}$ for $\chi_1$, and $\alpha=1.38^{+0.51}_{-0.33}$ and $\beta=1.61^{+0.64}_{-0.43}$ for $\chi_2$. We present the corresponding 2-dimensional confidence contours and 1-dimensional PDFs in Figure \ref{fig:contour_real} for comparison. These results of MCMC method demonstrate consistency with that of DNN within the 1$\sigma$ confidence level. On the other hand, it is discernible that the results obtained with MC samples diverge from the predictions based on central spin data within the $1\sigma$ confidence interval, attributed to the significant uncertainties inherent in spin observations. Our results demonstrate that it is not sufficient to fit the population model only with the central values of the spin observations. Furthermore, we also plot the distributions of the component spin magnitudes of BBHs within the parameters inferred from the DNN in Figure \ref{fig:contour_real_bands}. Notably, the distributions of spin magnitudes exhibit peaks at $\chi_1=0.30^{+0.05}_{-0.07}$ and $\chi_2=0.38^{+0.05}_{-0.02}$, seemingly suggesting the presence of non-zero spin for both black holes involved in the BBH merger.

\begin{figure*}[htbp]
 \centering
 \includegraphics[width=0.45\textwidth]{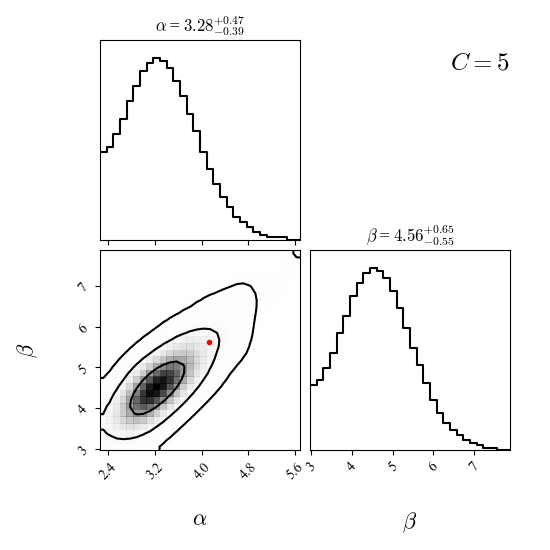}
 \includegraphics[width=0.45\textwidth]{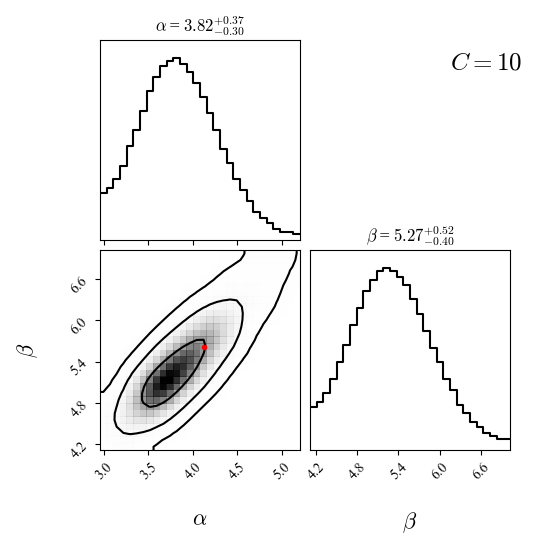}
 \caption{The 2-dimensional confidence contours and 1-dimensional PDFs for mock samples whose uncertainties are reduced by 5 (left) and 10 (right). The red points represent the parameters predicted with the observed central spins.}\label{fig:contour_real_reduce}
\end{figure*}

To investigate the relationship between observed uncertainty and population parameter predictions, we reduce the uncertainty of observation to simulate data and constrain the parameters. In the context of simulation, our focus is directed solely on one spin sample, i.e. the spin sample of the larger BH. Mock samples are generated via MC-bootstrap, sampling from a Gaussian distribution $G(\chi,\Delta_{\chi}/C)$, where $\chi$ denotes the observed central value of the spin sample of the larger BH, $\Delta_{\chi}$ represents the observation error, and $C$ signifies the coefficient by which the error is reduced. Subsequently, mock samples I and II are generated with $C=5$ and $C=10$, respectively, and the corresponding parameters are predicted using our DNN. The results are delineated in Figure \ref{fig:contour_real_reduce}, where the parameters predicted based on observed central spins are denoted by red points in the 2-dimensional confidence contours. Notably, the prediction of mock sample I ($C=5$) exhibits a slight deviation from the results of the central spin data prediction within the $1\sigma$ confidence level, while the prediction of mock sample II ($C=10$) aligns with it. These comparisons indicate that high-precision spin measurements can be directly used to rapidly obtain group parameter information in the future.

\begin{figure*}[htbp]
 \centering
 \includegraphics[width=0.3\textwidth]{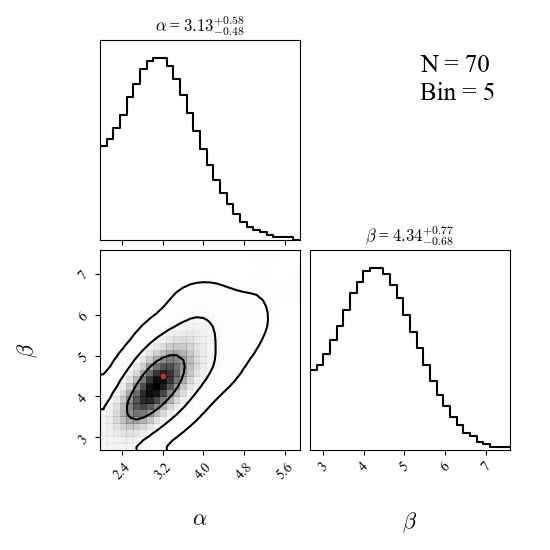}
 \includegraphics[width=0.3\textwidth]{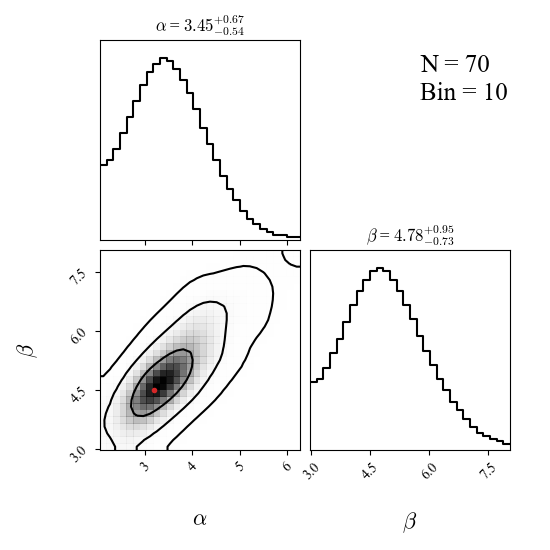}
 \includegraphics[width=0.3\textwidth]{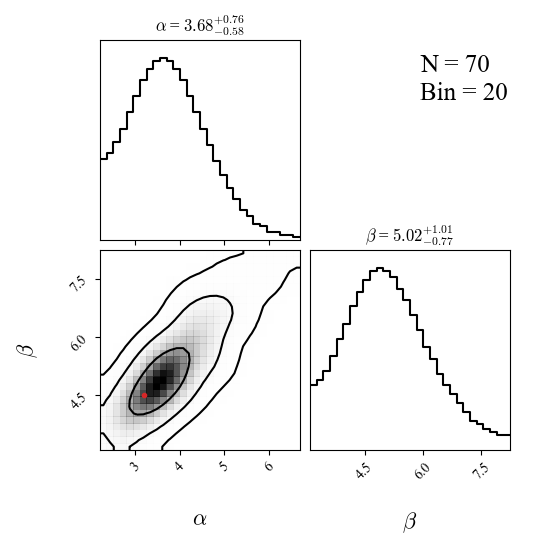}
 \includegraphics[width=0.3\textwidth]{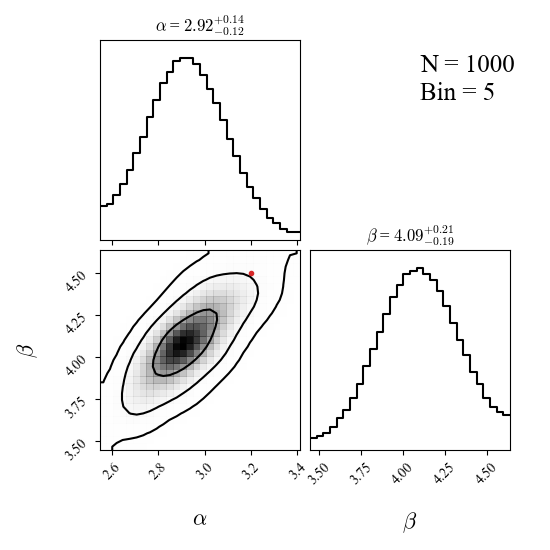}
 \includegraphics[width=0.3\textwidth]{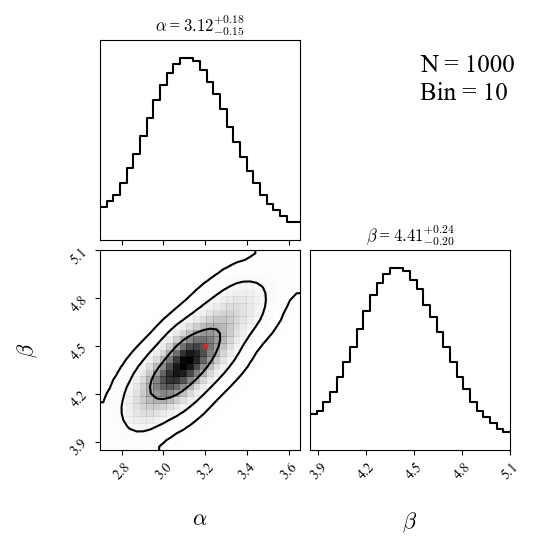}
 \includegraphics[width=0.3\textwidth]{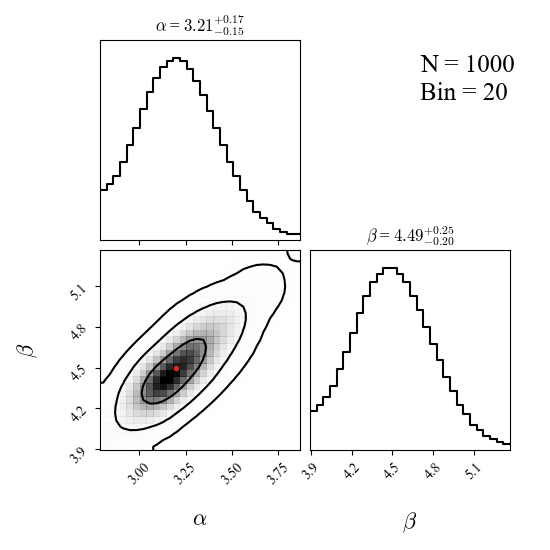}
 \caption{The 2-dimensional confidence contours and PDFs of parameters of samples with 70 and 1000 data points. }\label{fig:contour_N_B}
\end{figure*}

Given the anticipated surge in GW detections, projected to reach 10$^4$--10$^5$ events annually, owing to the heightened sensitivity of third-generation GW detectors such as the Einstein Telescope \cite{Ding:2015uha}, we undertake a comprehensive investigation into the efficacy of our DNN across varying data volumes, encompassing both present and projected future scenarios. Leveraging a fiducial spin distribution characterized by parameters $\alpha=3.2$ and $\beta=4.5$, we generate two distinct samples comprising 70 and 1000 data points, representative of current and anticipated future data volumes, respectively. By partitioning each sample into 10 bins and inputting counts into our well-trained DNN, we derive the predicted parameters. This procedure is iterated 1000 times to procure corresponding parameter chains. Additionally, we explore the influence of varying bin numbers, i.e. 5, 10, and 20, on parameter estimation. The results are presented in Table \ref{tab:par_N_B} and Figure \ref{fig:contour_N_B}, wherein the fiducial parameters are denoted by red points in the 2-dimensional confidence contours of parameters.

\begin{table}[htbp]
\centering                  
\caption{\small{The parameters of samples with 70 and 1000 data points.}} 
\arrayrulewidth=1.0pt          
\renewcommand{\arraystretch}{1.3}   
{\begin{tabular}{lcccc} 
\hline\hline 
& \multicolumn{2}{c}{70}& \multicolumn{2}{c}{1000}\\
\cline{2-5}
& $\alpha$ & $\beta$ &$\alpha$ & $\beta$\\ \hline
5	&$3.13^{+0.58}_{-0.48}$	&$4.34^{+0.77}_{-0.68}$ &$2.92^{+0.14}_{-0.12}$ &$4.09^{+0.21}_{-0.19}$\\
10	&$3.45^{+0.67}_{-0.54}$	&$4.78^{+0.95}_{-0.73}$ &$3.12^{+0.18}_{-0.15}$ &$4.41^{+0.24}_{-0.20}$\\
20	&$3.68^{+0.76}_{-0.58}$	&$5.02^{+1.01}_{-0.77}$ &$3.21^{+0.17}_{-0.15}$ &$4.49^{+0.25}_{-0.20}$\\
\hline
\end{tabular}}\label{tab:par_N_B}
\end{table}

Evidently, within the context of 70 data points, the predicted parameters consistently align with the fiducial values within the 1$\sigma$ confidence level across all bin numbers. For the dataset comprising 1000 data points, a deviation from the fiducial values of approximately 2$\sigma$ confidence level is observed when utilizing 5 bins. However, congruence with the fiducial values within the 1$\sigma$ confidence level is achieved for both 10 and 20 bins. Our results underscore the importance of employing an adequate number of bins to effectively capture the distribution characteristics, particularly for larger datasets. In our work, the utilization of 10 bins within the DNN framework is deemed suitable for both current and anticipated future data volumes, facilitating robust characterization of spin distributions across varying sample sizes.

\section{Conclusion and Discussion}\label{sec:summary}

Spin encapsulates pivotal information regarding the formation and evolution of black holes. Numerous studies have delved into the spin dynamics of BBH systems using GW observations. However, the discourse surrounding the spin characteristics of BH involved in BBH mergers remains contentious. In this work, we emploied a deep learning (DL) approach to probe the spin distribution of BBH systems within the GWTC-3 catalog. We devised a fully connected Deep Neural Network (DNN) tailored to extract population parameters pertaining to spin, leveraging the framework of the Beta distribution. While DL methods have been previously proposed for investigating the GW population \cite{Mould:2022ccw}, our approach diverges from their work. Mould et al. \cite{Mould:2022ccw} amalgamated DL methodologies with hierarchical Bayesian inference to explore the GW population landscape. In their study, DL techniques were utilized to reconstruct population models encompassing mass, spin, selection functions, and branching functions across merger generations. In contrast, our methodology is singularly focused on the spin model, inferring population parameters solely based on observational data. Eschewing hierarchical Bayesian techniques, our parameter constraint process is expedited. Furthermore, our approach of segregating data into 10 bins affords flexibility in accommodating GW event datasets of varying sizes in the future. In order to determine the parameters and associated uncertainties, we employed Monte Carlo-bootstrap sampling to generate samples for parameter prediction using the DNN, subsequently assessing uncertainty through 1000 iterations of sampling. Within the spin sample of the larger BH, the parameters characterizing the spin distribution were constrained to $\alpha=1.30^{+0.25}_{-0.18}$ and $\beta=1.70^{+0.42}_{-0.29}$. Within spin sample of the smaller BH, parameters were $\alpha=1.37^{+0.31}_{-0.20}$ and $\beta=1.63^{+0.30}_{-0.20}$. Notably, these parameters deviate from the results with the central values of the spin data confidence level. This deviation is attributed to the substantial uncertainty associated with spin observations, which hindered the generation of samples closely resembling the central values. Upon reducing the spin uncertainty by a factor of 10, the parameters aligned with the central value predictions within the 1$\sigma$ confidence level. Additionally, we also compared the parameters of spin model from DL and MCMC method, and found they are consistent with each other within 1$\sigma$ confidence level.

The spin distribution has been extensively investigated using various methodologies, primarily based on Hierarchical Bayesian inference techniques, resulting in differing conclusions \cite{KAGRA:2021duu,LIGOScientific:2018jsj,Galaudage:2021rkt,LIGOScientific:2020kqk,Callister:2022qwb,Golomb:2022bon}. For instance, Abbott et al. \cite{LIGOScientific:2018jsj} considered the degeneracy between mass and spin, modeling these distributions together using a Beta distribution. Their analysis of 10 BBH mergers revealed that the spin distribution peaks at approximately 0.2, with 90$\%$ of BH spins less than 0.55 and 50$\%$ less than 0.27. Similar results were observed in the Beta model using the GWTC-2 data \cite{Galaudage:2021rkt}. Golomb and Talbot \cite{Golomb:2022bon}, without making strong assumptions about the spin shape, used flexible cubic spline interpolants to model the spin observations of 59 events in the third observing run (O3) of the LIGO-Virgo network, finding that 77.1$\%$ of BHs have spin magnitudes less than 0.5, with 50$\%$ below 0.25. The proportion of high spin magnitudes appears to increase with the number of detected GW events. However, in the GWTC-3 analysis, Abbott et al. \cite{KAGRA:2021duu} reported a peak at 0.13 with a tail extending toward larger values, and 50$\%$ of BHs have spin magnitudes below 0.25. In this work, we modeled the spin observations of GWTC-3 with a Beta distribution and found that the distributions of spin magnitudes exhibit peaks at $\chi_1=0.30^{+0.05}_{-0.07}$ and $\chi_2=0.38^{+0.05}_{-0.02}$. Specifically, we observed that in the $\chi_1$ distribution, 61$\%$ of BHs have spin magnitudes less than 0.5, with 50$\%$ below 0.41. In the $\chi_2$ distribution, 56$\%$ have spin magnitudes less than 0.5, with 50$\%$ below 0.45, suggesting a higher spin magnitude compared to previous studies. These differences highlight the impact of varied methodologies, data samples, and spin models on the inferred spin distributions. Unlike previous studies, our methodology did not incorporate a broad range of factors, such as merger rates and selection effects. Additionally, our data analysis focused exclusively on spin, without accounting for variables such as mass and spin tilt angle.

In conclusion, our findings support the notion that the spin distribution adheres to a Beta model, exhibiting a peak around $\sim$0.35, suggesting the likelihood of both black holes in the BBH merger possessing non-zero spins. BHs with non-zero spins may stem from multi-generational mergers within hierarchical black-hole merger scenarios \cite{Gerosa:2021mno,Callister:2022qwb}. As two BHs merge, the spin of the resulting BH reflects a combination of the spins of the merging BHs and the orbital angular momentum at plunge. In hierarchical mergers, prevalent in star clusters and active galactic nuclei (AGNs), the spin of BHs resulting from mergers involving second or higher-generation BHs may be significant compared to first-generation BHs originating from stellar collapse. Moreover, if the spins of the binary BHs are aligned with the orbital angular momentum, the resultant remnant spin is likely to be larger.

It is noteworthy that our study represents an exploratory endeavor aimed at examining spin distribution through the utilization of deep learning methodologies, specifically relying on existing GW observations from ground-based detectors. Consequently, certain factors, such as selection effects, were not taken into account in our analysis. The primary objective of our investigation is to furnish a methodological framework conducive to analyzing forthcoming, substantially larger GW datasets, thereby facilitating a comprehensive understanding of BH characteristics within our cosmic milieu. This aspiration is anticipated to be realized with the advent of both ground-based and space-based gravitational wave detectors. Furthermore, our study refrains from comparing distinct spin models utilizing our DNN. Nevertheless, our DNN architecture holds potential as a modular template, serving as a foundation for constructing more intricate DNNs capable of comparing and inferring various spin models in future research endeavors.

\vspace{5mm}
\centerline{\rule{80mm}{0.5pt}}
\vspace{2mm}

\bibliographystyle{cpc}
\bibliography{reference}

\end{document}